\documentclass{emulateapj}
\usepackage{psfig,enumerate}

\newcommand{\dgr}{$^{\circ}~$}

\newcommand{\h}{$^{\rm h}$}
\newcommand{\m}{$^{\rm m}$}

\newcommand{\se}{$^{\rm s}$}

\newcommand{\hi}{H{\footnotesize I} }

\begin{document}

\title{Discovery of a Large Stellar Periphery Around the Small Magellanic Cloud}

\shorttitle{The SMC Stellar Periphery}
\shortauthors{NIDEVER ET AL.}

\author{David L. Nidever\altaffilmark{1},
Steven R. Majewski\altaffilmark{1}, 
Ricardo R. Mu\~noz\altaffilmark{2}, \\
Rachael L. Beaton\altaffilmark{1},
Richard J. Patterson\altaffilmark{1},
\& William E. Kunkel\altaffilmark{3}
}

\altaffiltext{1}{Dept. of Astronomy, University of Virginia,
Charlottesville, VA, 22904-4325, USA (dnidever, srm4n, rlb9n, rjp0i@virginia.edu)}

\altaffiltext{2}{Departamento de Astronom\'ia, Universidad de Chile, Casilla 36-D,
Santiago, Chile (rmunoz@das.uchile.cl)}

\altaffiltext{3}{Las Campanas Observatory, Carnegie Institution of Washington,
Casilla 601, La Serena, Chile (kunkel@lco.cl)}

\begin{abstract}
The Magellanic Clouds are a local laboratory for understanding the
evolution and properties of dwarf irregular galaxies.  To reveal the
extended structure and interaction history of the Magellanic Clouds we
have undertaken a large-scale photometric and spectroscopic study of
their stellar periphery (the MAgellanic Periphery Survey, MAPS).  We
present first MAPS results for the Small Magellanic Cloud (SMC):
Washington $M$, $T_2$ + DDO51 photometry reveals metal-poor red giant branch
stars in the SMC that extend to large radii ($\sim$11 kpc), are
distributed nearly azimuthally symmetrically (ellipticity=0.1), and
are well-fitted by an exponential profile (out to $R$$\approx$7.5\degr).
An $\sim6$ Gyr old, [Fe/H]$\approx$$-$1.3 main-sequence turnoff is also
evident to at least $R$=7.3\degr, and as far as 8.4\dgr in some directions.
We find evidence for a ``break'' population beyond  $\sim$8 radial scalelengths having a very
shallow radial density profile that could be either a bound stellar halo or
a population of extratidal stars.
The distribution of the intermediate stellar component (3 $\lesssim$ $R$ $\lesssim$ 7.5\degr) contrasts with
that of the inner stellar component ($R$ $\lesssim$ 3\degr),
which is both more elliptical ($\epsilon$ $\approx$ 0.3) and offset from the center of the intermediate
component by 0.59\degr, although both components share a similar radial
exponential scale length.
This offset is likely due to a perspective effect because stars on the
eastern side of the SMC are closer on average than stars
on the western side.  This mapping of its outer stellar
structures indicates that the SMC is more complex than previously
thought.
\end{abstract}

\keywords{Galaxies: interactions --- Local Group --- Magellanic Clouds
--- Galaxies: dwarf --- Galaxies: individual (SMC) --- Galaxies: photometry }

\section{Introduction}
\label{sec:intro}

The Magellanic Clouds (MCs) have had a complex history that includes a
recent close encounter with each other (about $\sim$200 Myr ago), a
presently ongoing period of intense star formation, and immense loss
of gas to the various components of the \hi Magellanic System (the
Magellanic Bridge, Magellanic Stream and Leading Arm).  Systematic
study of the Small Magellanic Cloud (SMC) stellar morphology started
with the pioneering work by \citet{deV55}, who studied mostly the
bright and young structures and noted the SMC's very irregular shape.
Photographic plate photometry of red horizontal branch stars (RHB)
helped show that the older SMC population extends to a radius of
$\sim$5\degr, is fairly azimuthally symmetric, but has a larger
line-of-sight depth on the northeastern side than for the rest of the
SMC likely due to the recent interaction with the Large Magellanic
Cloud \citep[LMC;][]{HH89,GH91,GH92}.  Additionally, \citet{GH91}
found that the SMC line-of-sight depth on the western side increases
with radius, a finding more consistent with a spherically symmetric
radial exponential law than with a flattened system (i.e.,~a disk).
Stellar radial velocity studies \citep{Kunkel00,HZ06} show no signs of
rotation but instead that the stars are pressure supported and very
likely in a spherical or ellipsoidal structure. In contrast, rotation
is observed in the \hi component of the SMC for $R\lesssim$3\degr
\citep{Stani04}, which reinforces that the structure of the SMC is
complex and multi-faceted.  In deep CCD photometry of three fields to
the south of the SMC, \citet{NG07} found old SMC stars out to
5.8\degr, whereas \citet{DePropris10} find very few SMC stars at a
radius of $\sim$5\dgr toward the east and suggest the edge of the SMC
is at a radius of 6 kpc.  All of the above results for the SMC are
generally in agreement with the following known properties of dIrr
galaxies: the old and red stellar components (as traced by evolved
stars) are (1) spatially smoother and more extended than their younger
and bluer counterparts, and (2) generally follow a radial exponential
profile \citep[e.g.,][]{Mateo98}.  However, many questions remain
about the structure of the SMC periphery beyond $R$$\sim$5\degr: Where
is the edge of the SMC?  Does the SMC have a stellar halo or stellar
tidal tails?  How has the interaction of the MCs with each other (and
the MW) disturbed their stellar structures?

We have undertaken the MAgellanic Periphery Survey (MAPS), a large
photometric and spectroscopic survey of the stellar periphery of the
MCs designed to elucidate their extended structure and kinematics and
piece together their past interactions as recorded in their outer
morphologies.  In this Letter, we present our first photometric
results for the SMC (\S 2).  Using Washington $M$, $T_2$ + DDO51
photometry we detect metal-poor SMC red giant branch stars to a radius
of 10.6\dgr (11.1 kpc) and at all of the many position angles explored
around the SMC (\S 3).  These stars are distributed nearly azimuthally
symmetrically and are well-fitted by a radial exponential profile to
$R$ $\approx$ 7.5\degr.  We find evidence for a ``break'' population
beyond $R$ $\sim$ 8\dgr with a shallower radial density profile that
could be a bound stellar halo or a population of extratidal stars (\$
4). The fitted center of the intermediate stellar distribution
(3$\lesssim$ $R$ $\lesssim$7.5\degr) is offset by 0.59\dgr from that
of the inner stellar distribution, which is also much more elliptical
This lack of centration is likely due to a perspective effect (\S 4).

\section{Observations And Data Reduction}
\label{sec:red}

\begin{figure}[t]
\begin{center}
\fbox{\includegraphics[angle=0,scale=0.45]{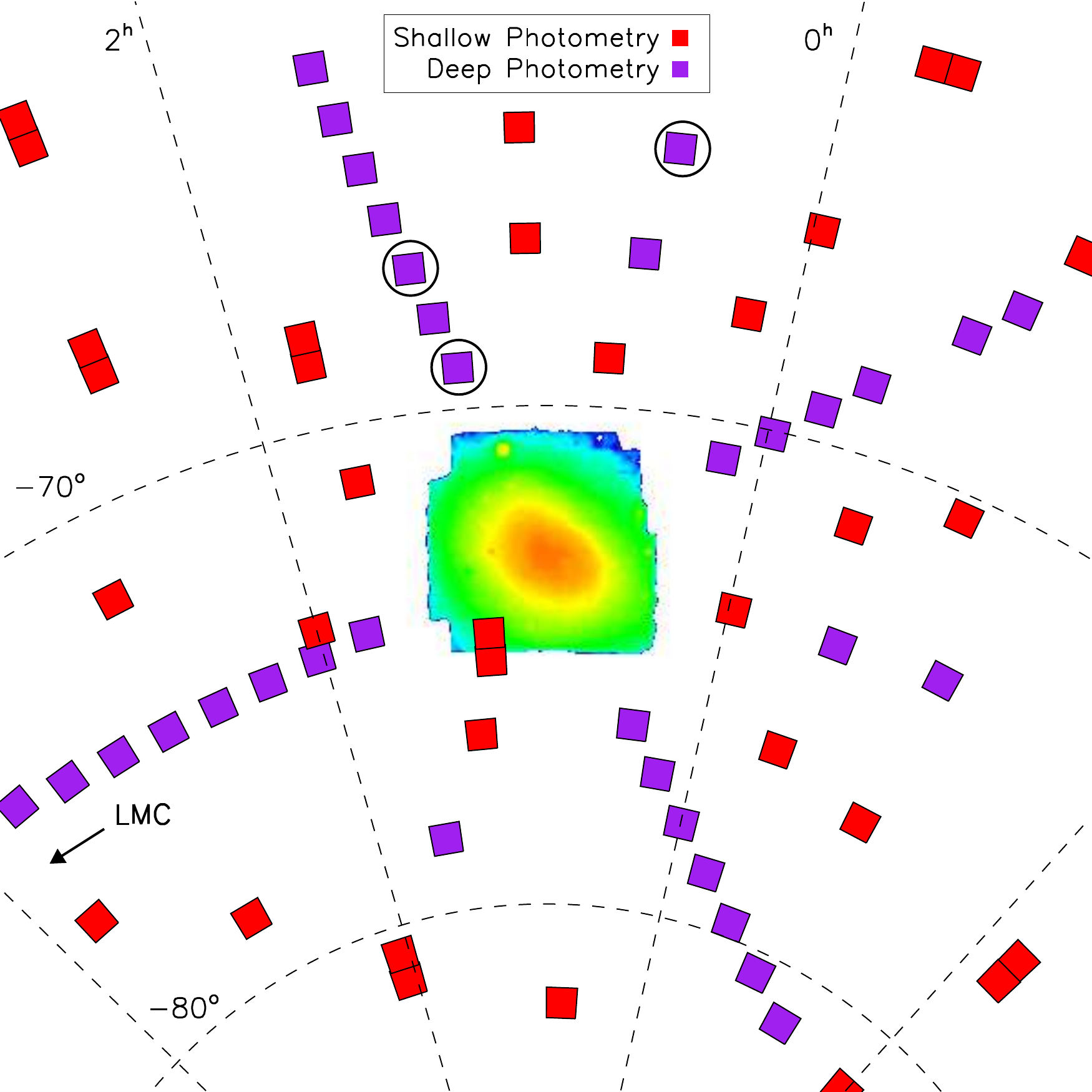}}
\end{center}
\caption{MAPS fields of the SMC periphery.  Shallow fields are in red ($M$$\lesssim$22) while deeper fields are in purple
($M$$\lesssim$24).  The central image shows the RGB starcounts from OGLE-III and MCPS.  Color magnitude
diagrams and color-color diagrams for the circled fields are shown in Fig.\ \ref{fig_cmd_sequence}.}
\label{fig_mapsfields}
\end{figure}

The stellar periphery of the SMC was imaged as part of the MAPS survey
using the Washington $M, T_2 +$ DDO51 photometric system
\citep{Majewski00}, which substantially helps discriminate foreground
MW dwarf stars from the Magellanic red giant branch (RGB) stars.
Observations were obtained with the MOSAIC II camera
(36\arcmin$\times$36\arcmin) on the CTIO--4m Blanco telescope on UT
2006 February 28--March 4 and UT 2007 August 16--18 with the majority
of the SMC data coming from the second observing run.  In all fields
net integrations of (40, 60, 280)s in ($M$, $T_2$, DDO51) were
acquired, and in a subset of 33 fields (most arranged in four radial
``spokes'' around the SMC) additional 300s observations were also
obtained in $M$ and $T_2$ (see Fig.\ \ref{fig_mapsfields}).
Observations were also obtained for \citet{Geisler90} standard star
fields to enable photometric calibration.

The SuperMACHO pipeline \citep{Rest05,Miknaitis07} was used to obtain
flattened images and the PHOTRED pipeline (D. Nidever et al.\ 2011, in
preparation) was used for the rest of the photometric reduction.
PHOTRED makes use of a combination of the stand-alone DAOPHOT/ALLSTAR
\citep{Stetson87}, ALLFRAME \citep{Stetson94} and Sextractor
\citep{Bertin96} packages to make photometric measurements and create
final, calibrated, dereddened photometry.  Figure \ref{fig_mapsfields}
shows our MAPS SMC fields and the density of SMC RGB stars in the
central region ($R$$\lesssim$2\degr; color image) as selected from the
OGLE-III \citep{ogle3} and MCPS \citep{ZH02} SMC photometric catalogs.
Examples of the final MAPS SMC photometry can be seen in Figure
\ref{fig_cmd_sequence}.

\section{The Radial Density Profile}
\label{sec:profile}

\begin{figure*}[t]
\includegraphics[angle=0,scale=0.48]{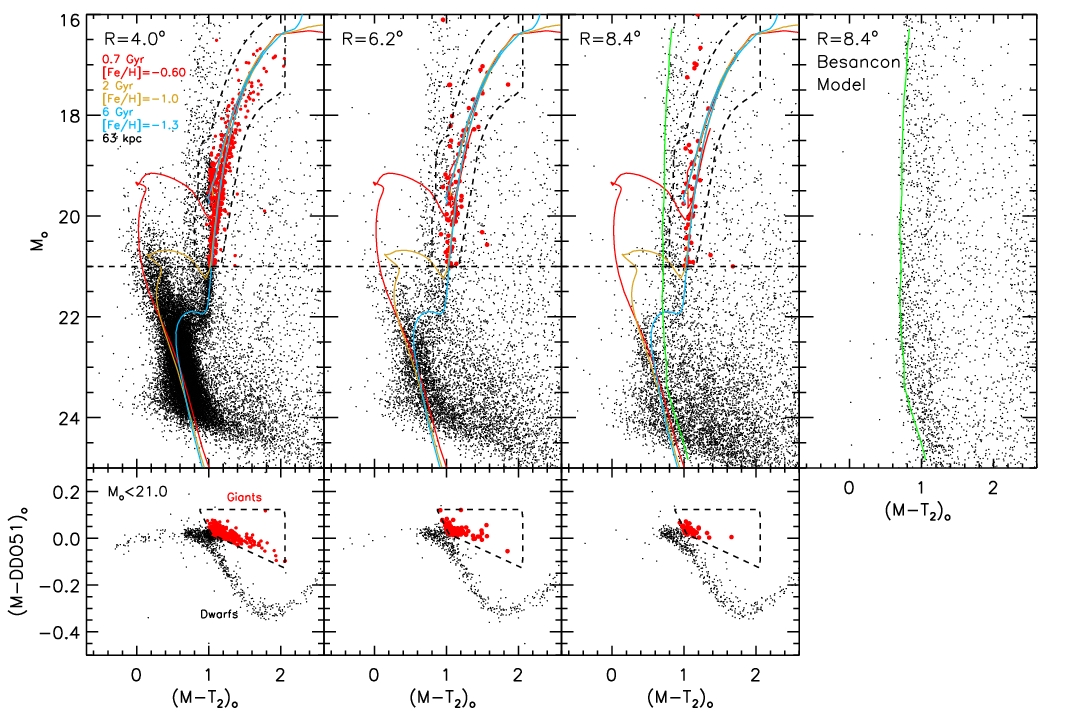}
\caption{
Photometric data for three ``deep'' SMC MAPS fields at
$R$=4.0, 6.2 and 8.4\dgr (circled in Fig.\ \ref{fig_mapsfields}).  The top panels show \citet{SFD98}
dereddened  ($M$-$T_2$,$M$) CMDs with some fiducial Padova isochrones shifted to 63 kpc
([Fe/H]$=-$0.60/0.7 Gyr in red; [Fe/H]$=-$1.0/2 Gyr in gold; and [Fe/H]$=-$1.3/6 Gyr in blue).
The dashed line shows the region used to select SMC RGB stars.
The bottom panels show dereddened ($M$-$T_2$, $M$-DDO51)
2CDs for $M_0\leq$21.0 with the giant selection (dashed line).
Giants are plotted with filled red circles in both top and bottom panels.
SMC giants and MS stars are visible in all three fields.  The
right-most panel shows simulated photometry for the $R$$=$8.4\dgr
field using the Besancon Galactic model to ascertain the approximate
distribution of MW foreground stars, which are roughly bounded (on the blue edge) by
the green line (also shown in the $R$=$8.4$\dgr panel).}
\label{fig_cmd_sequence}
\end{figure*}

Figure \ref{fig_cmd_sequence} shows dereddened ($M$$-$$T_2$, $M$)
color magnitude diagrams (CMDs) and ($M$$-$$T_2$, $M$$-$DDO51)
color-color diagrams (2CDs) for three fields in the northern part of
the SMC.  These fields are typical for the north and northeastern SMC,
which have higher densities than in other directions as described
below.  The SMC main-sequence (MS), RGB, and red clump (RC) are
visible in the three fields extending out to $R$$=$8.4\degr.  The
right-most panel is simulated photometry for the $R$$=$8.4\dgr field
using the Besancon Galactic model \citep{Robin03} to ascertain the
approximate distribution of MW foreground stars.  Padova isochrones
\citep{Girardi02} with [Fe/H]=$-$0.6/0.7 Gyr, [Fe/H]=$-$1.0/2 Gyr, and
[Fe/H]=$-$1.3/6 Gyr, shifted to a distance of 63 kpc, are also shown
(red, gold and blue respectively) as fiducials.  The dominant
population in the left three panels is $\sim$6 Gyr old but a sizable
younger population ($\sim$2 Gyr) is also visible to at least
$R$=6.2\dgr as well as a weaker $\sim$0.7 Gyr population.  Giant and
dwarf stars separate well in the 2CD due to the gravity sensitivity of
the DDO51 filter.  Our giant selection criteria are shown by the
dashed line and all stars selected as giants (with $M_0$$<$21.0) are
plotted as filled red circles.

We derive starcounts in our fields by using a 2CD giant selection
(dashed line in Fig.\ \ref{fig_cmd_sequence} lower panels) and an SMC
RGB-CMD selection (dashed line in Fig.\ \ref{fig_cmd_sequence} upper
panels).  Our starcounts can be contaminated by MW halo giants and
metal-poor MW dwarfs that fall in the 2CD giant selection.  To
estimate and remove this contamination we use the same 2CD selection
criteria and the same CMD selection criteria but offset by $-$3
magnitudes in order to not overlap the SMC stars (following Majewski
et al.\ 2000b; see, e.g., their Fig.\ 12).  This shifted CMD selection
extends beyond our bright limit; to account for missing giants we use
a related dataset \citep{Majewski04} with the same photometric filters
and reduced with the same software but that extends to brighter
magnitudes in fields far enough from the MCs not to be contaminated by
them; from these we calculate that we are on average missing 22\% of
the foreground giants due to our brightness limit\footnote{This
fraction is small because in general the RGB-tip is less well
populated.}.  This constant fraction is used to correct the foreground
values for all fields.  The selection at these brighter magnitudes
should have contamination levels similar to those in our actual SMC
selection if the Galactic halo field giants follow a $\sim$$R^{-3}$
law, which produces a uniform distribution in magnitude
\citep{Carina00}.  This method has been used (and more extensively
explained) in several related studies \citep{Carina00, Palma03,
Westfall06} and shown to be reliable \citep{Majewski05}.  Finally, we
perform a 2D spatial linear fit to the foreground contamination levels
of all fields to obtain a smooth distribution.

\begin{figure*}[t]
\begin{center}
$\begin{array}{cc}
\includegraphics[angle=0,scale=0.38]{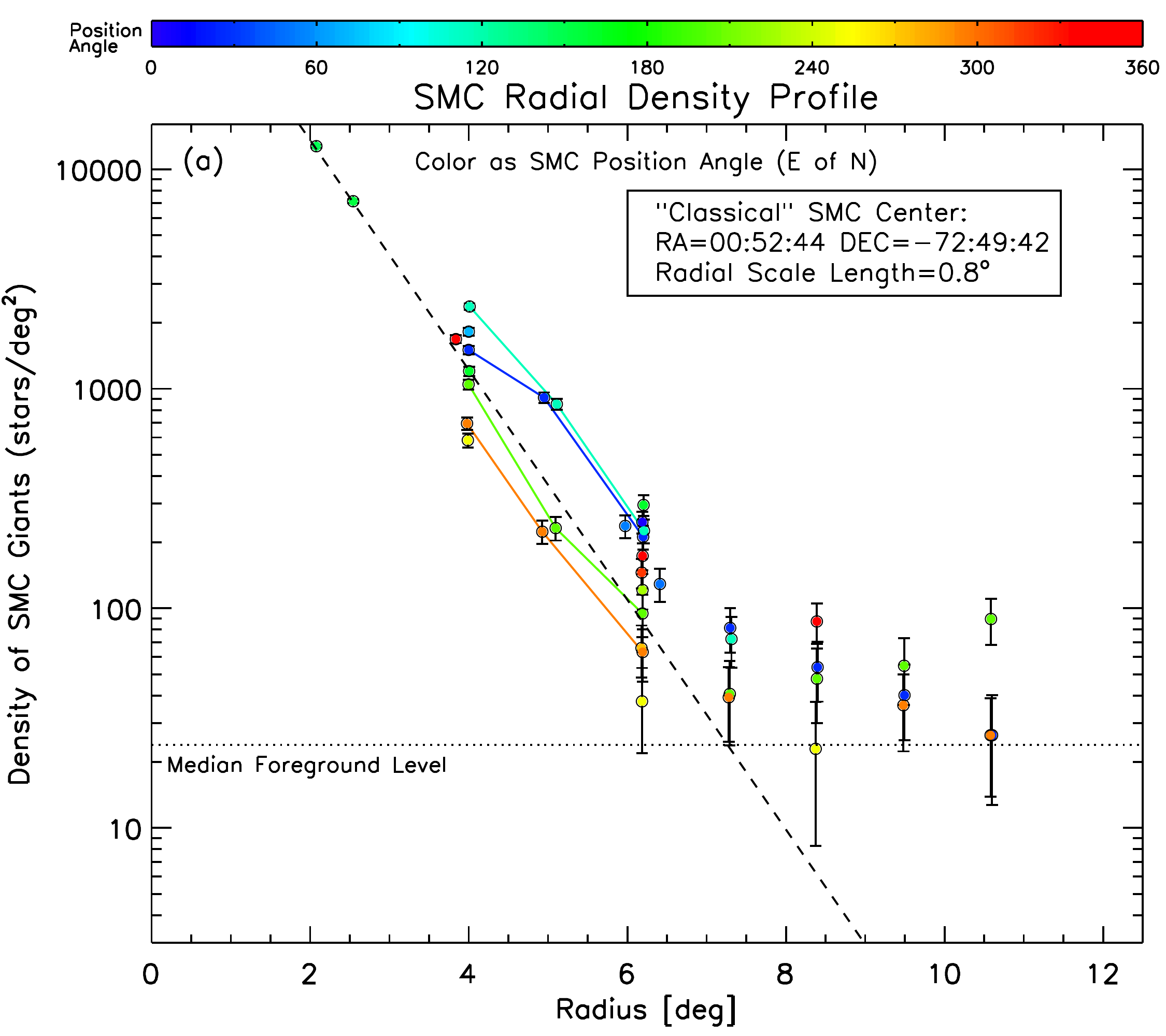} &
\includegraphics[angle=0,scale=0.38]{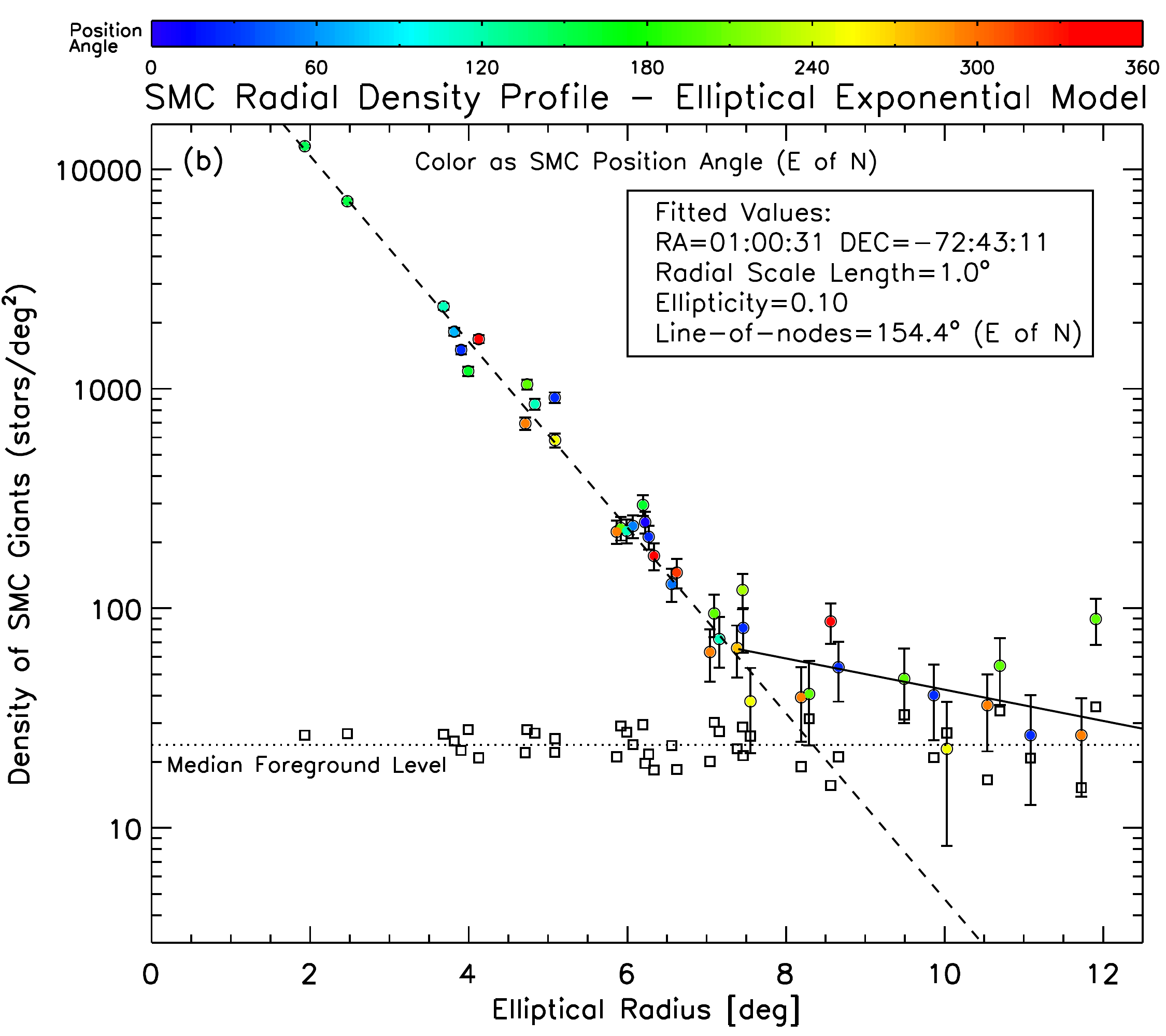}
\end{array}$
\end{center}
\caption{The SMC radial density profile. ({\em a}) Giant starcounts
from our MAPS fields as a function of radial distance from the
\citet{Mateo98} SMC center.  Color indicates the position angle (PA)
of the fields (East of North; PA=116\dgr is towards the LMC) and lines
connect the fields of the four ``radial spokes'' (for the inner three
fields).  The best-fit exponential profile (to the $R<$7\dgr fields)
with $h_{\rm r}$=0.8\dgr is shown by the dashed line.  The dotted line
indicates the median foreground level.  Large offsets in the
starcounts are apparent and indicate that the center used is not
appropriate for the SMC periphery.  ({\em b}) Giant starcounts as a
function of major axis radius for our best-fit elliptical exponential
model.  This profile is a much better fit to all of the data than the
model in ({\em a}).  A ``break'' population at $R$$>$7.5\dgr is
apparent.  The estimated foreground contamination levels for each
field are shown as open squares.}
\label{fig_radprofile}
\end{figure*}

The starcounts as a function of radial distance from the SMC center
\citep[taken as $\alpha$$=$00\h 52\m 44\se, $\delta$$=$$-$72\degr
49.7\arcmin;][]{Mateo98} can be seen in Figure \ref{fig_radprofile}a.
The median foreground level is shown.  Fields with $R>7.5$\dgr towards
the LMC (four fields) were removed because these are heavily
``contaminated'' by LMC RGB stars.  We detect SMC stars above the
foreground in all fields to $R$$=$10.6\dgr (11.1 kpc), almost twice as
far as the previous most distant photometric detection of SMC stars
\citep[$R$ $\sim$ 5.8\degr,][]{NG07}.

\begin{figure}[ht!]
\begin{center}
\fbox{\includegraphics[angle=0,scale=0.45]{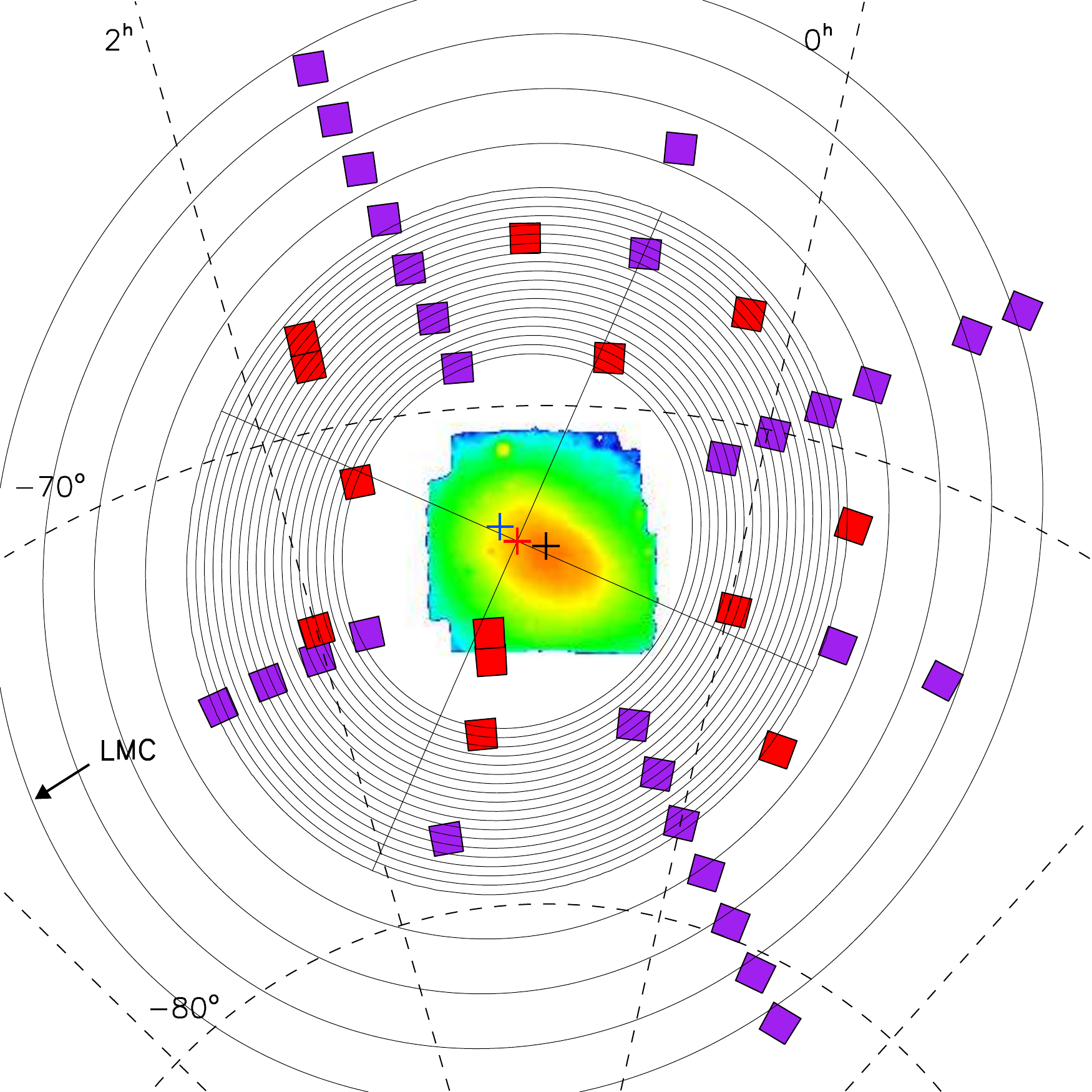}}
\end{center}
\caption{Stellar surface density of the SMC.  The central colored
image shows the RGB starcounts from OGLE-III and MCPS.  Contours show
our best-fit model to the SMC RGB starcounts in MAPS fields and are
drawn at even intervals in log density (starting at 33 giants
deg$^{-2}$ and increasing with an interval of 0.08 in log density).
Squares show MAPS fields constraining the model fit.  The optical
center (black cross), HI dynamical center \citep[][blue
cross]{Stani04}, and model center (red cross) are indicated.  }
\label{fig_mapmodel}
\end{figure}

The SMC starcount density follows a general radial exponential, but
the distribution is not well-fitted by a single exponential for all
radii and position angles, as can be seen by the large scatter about
the exponential fit (dashed line, fit to $R$ $<$ 8\dgr fields).  The
density along each radial spoke follows roughly a radial exponential
profile with nearly identical radial scale lengths ($h_{\rm r}$
$\approx$ 0.8\degr) but with different amplitudes -- northeastern
fields having higher amplitudes than southwestern fields.  A simple
explanation of this pattern is that the center we have used -- the
center of the inner stellar distribution -- is not appropriate for the
outer SMC stellar distribution.  Figure \ref{fig_radprofile}b shows an
elliptical exponential fit to the same fields that also allowed the
density center to vary.  This provides a much better solution than the
previous fit, with best-fitting parameters: $\alpha_{\rm center}$$=$01\h 00\m 31\se,
$\delta_{\rm center}$=$-$72\degr 43\arcmin
11\arcsec $\pm$4\arcmin~(0.59\dgr from the Mateo 1998 center),
$h_{\rm r}$$=$1.0\degr, ellipticity$=$0.10, and line-of-nodes$=$154.4\dgr (E
of N).  An illustration of the fitted distribution is shown in Figure
\ref{fig_mapmodel}.  The shape of the outer contours ($R$$>$7.5\degr)
is not well constrained by our data since we only sample a few
position angles.  For Figure \ref{fig_mapmodel} the ellipticity and
line-of-nodes of the intermediate distribution were simply extended to
the outer distribution.

The distribution of RGB stars is quite different for the inner
($R$$\lesssim$3\degr) and intermediate
(3$\lesssim$$R$$\lesssim$7.5\degr) radial regions.  The intermediate
component is much more azimuthally symmetric ($\epsilon$$=$0.1) than
the inner component, which is quite elliptical
\citep[$\epsilon$$=$0.3;][]{HZ06}, and the major axes of the two
components are nearly perpendicular \citep[difference of
$\sim$105\degr; inner line-of-nodes $\approx$ 50\degr;][]{HZ06}.  The
center of the intermediate component is offset by 0.59\dgr to the
northeast of the center of the inner component (red and black crosses
respectively in Fig.\ \ref{fig_mapmodel}) -- along the inner's major
axis -- and is closer to the \hi dynamical center
\citep[separation=0.38\degr;][]{Stani04}.

Finally, Figure \ref{fig_radprofile}b, and the sudden increase in
contour spacing seen in the outer parts of Figure \ref{fig_mapmodel},
both show that there is a striking departure from the exponential fit
for fields at large radii ($R$$\gtrsim$7.5\degr).  This ``break''
population extends from $R$$\approx$7.5\dgr to our fields at
$R$=10.6\dgr and has a much shallower radial density profile
($h_{\rm r}$$\approx$7\degr).  Because the detected break population density is
only a few times the foreground density, more data are needed to
determine its density at these large radii securely.

\section{Discussion}
\label{sec:discussion}

It is common for dIrr galaxies to have patchy distributions of young,
blue stars but smoother and more extended distributions of older,
redder stars that follow exponential (or King) surface density
profiles \citep{Mateo98}.  The fairly azimuthally symmetric elliptical
exponential profile of the older SMC stars is consistent with this
general trend.  However, this symmetrical distribution
($\epsilon$$\approx$0.1) of the stellar periphery is quite surprising
in the specific case of the SMC considering (1) the SMC's recent close
encounter with the LMC $\sim$200 Myr ago, which stripped large amounts
of \hi from the SMC to create the Magellanic Bridge
\citep[e.g.,][]{MB07}, (2) that the MCs have just recently passed
periGalacticon \citep[e.g.,][]{Besla07} and thus were subjected to
stronger tidal forces from the MW, and (3) that the LMC's outer
stellar contours are more elliptical
\citep[$\epsilon$$\approx$0.2;][]{vdM01} than the SMC's even though
the LMC is 10$\times$ more massive.  How is this possible?

On closer inspection, we find that the SMC is not as ``undisturbed''
as might be suggested from the fairly symmetrical density profile of
its intermediate component.  \citet{Hatz89} and \citet{GH91} used red
clump stars to show that the line-of-sight depth varies significantly
across the face of the SMC.  It is much thicker (``puffed-up'') in the
northeast and extends to smaller distances there (from the Sun)
compared to the southwest.  Therefore, the three-dimensional structure
of the SMC is somewhat irregular and disturbed, likely due to the
recent close interaction of the SMC with the LMC.  The fact that the
projected density profile is still so symmetric might be a coincidence
owing to the particular orientation of the MC system relative to our
viewing angle.

The difference in the line-of-sight depths between the eastern and
western sides of the SMC suggests a possible explanation for the lack
of centration between the inner and outer distributions.  Because the
stars on the eastern side of the SMC are {\em on average} closer than
the stars on the western side, the SMC periphery will exhibit a
perspective effect.  The near-side of the SMC will appear larger than
the far-side and the density contours will be ``stretched'' on the
near-side (east) and ``bunched-up'' on the far-side (west).  The net
result is that the center of outer contours of a fit to the on-sky
densities will be systematically shifted to the near-side, or the
east, as observed.  A perspective effect has previously been observed
to affect the outer contours of the LMC due to the inclination of its
stellar disk \citep{vdM01}.  However, it is important to note that the
lack of any stellar rotation seen in radial velocities
\citep{Kunkel00,HZ06} or proper motions \citep{Piatek08} indicates
that the SMC does not have a stellar disk (like the LMC), but is
pressure supported and likely has a more spheroidal or ellipsoidal
shape.  The difference in line-of-sight distances across the SMC might
be a result of tidal distortions of the SMC shape due to the recent
encounter of the MCs $\sim$200 Myr ago.

What is the nature of the break population that apparently dominates
over 7.5$\lesssim$$R$$\lesssim$10.6\dgr and is fairly azimuthally
symmetric for at least 270\degr?  The two main possibilities are (1) a
tidal tail or debris, or (2) a ``classical'' bound stellar halo as
exists around the MW and M31 \citep[e.g.,][]{raja05}.  The MCs had a
recent close encounter ($\sim$200 Myr ago), which no doubt had a
profound impact on the SMC.  The LMC very likely stripped \hi gas from
the SMC to form the Magellanic Bridge, and possibly also puffed-up the
stellar distribution on the northeastern side of the SMC.  It's
possible that this encounter also created a stellar tidal stream.
Extratidal stars energy sort into the well-known double-tidal tail
shape only outside a few tidal radii and at smaller radii can appear
fairly azimuthally symmetric \citep{Munoz08}.  Since the tidal radius
of the SMC is 4--9 kpc \citep{Stani04} the break population is within
$\sim$2 tidal radii; it is, therefore, difficult to distinguish a
bound classical halo from extratidal stars in this regime.  An anlysis
of the kinematics and velocity dispersion of these stars as well as
further photometric mapping at more position angles and larger radii
might help elucidate their true nature (i.e., bound versus unbound).
Either way, these stars can be thought of as a newfound ``halo''
component of the SMC.

\citet{DePropris10} claim that the edge of the SMC is at a radius of
$\sim$6 kpc on the eastern side because they detected only five
spectroscopically-confirmed SMC giants in an eastern field at
$R$$\sim$5 kpc.  We have a field $\sim$1.6\dgr away (but at a similar
radius) from theirs and find 57 giant candidates with our
Washington$+$DDO51 selection method ($M$$\lesssim$21.0).  When we also
use the De Propis et al.\ 2MASS selection criteria (13 $<$ $K_{\rm s}$
$<$ 14, 0.5 $<$ $J$$-$$K_{\rm s}$ $<$ 1.5) this reduces the number of
giants to five in agreement with the De Propris et al.\ result.  The
difference in detected giants between the two techniques is due to the
shallowness of the De Propis et al.\ selection, which is limited to
$M$$\sim$17 at the color of the SMC RGB ($M$$-$$T_2$$\sim$1.7).  As
can be seen in the $R$=6.2\dgr panel of Figure \ref{fig_cmd_sequence}
this selection only samples the very tip of the SMC RGB whereas our
selection goes four magnitudes fainter and is more sensitive to lower
SMC densities.  Therefore, we find the De Propris et al.\ conclusion
that the SMC edge toward the east is at $R$$\sim$6 kpc to be
premature; our deeper data show that the SMC extends to much larger
radii (at least $R$$\sim$9 kpc and likely to $R$$\sim$11 kpc, Fig.\
\ref{fig_radprofile}).

\section{Summary}
\label{sec:summary}

In this paper we have discovered SMC stars to a radius nearly twice as
large as the previous most-distant SMC detections.  We observe two
outer SMC stellar structures:
\begin{enumerate}[1.]
\item An intermediate component of older stars dominates over
3$\lesssim$$R$$\lesssim$7.5\degr, follows a slightly elliptical
exponential profile ($h_{\rm r}$=1.0\degr), is nearly azimuthally
symmetric ($\epsilon$=0.1), and has a center that is offset from the
center of the inner stellar distribution by 0.59\dgr (to the
northeast) likely due to perspective effects.  The structure of this
component is probably spheroidal or ellipsoidal.
\item An outer component dominating 7.5$\lesssim$$R$$\lesssim$10.6\dgr
that is fairly azimuthally symmetric over at least 270\dgr and has a
shallow radial scale length ($h_{\rm r}$$\approx$7\degr).  The
component could be a bound stellar halo or a population of extratidal
stars.  Analysis of kinematics and velocity dispersions of these stars
and further photometric mapping are needed to better reveal the nature
of this component.
\end{enumerate}
The SMC, therefore, appears to be much more complex than previously
thought and composed of several structural components, like larger
galaxies.

\acknowledgements

D.L.N. was supported by an Sloan Digital Sky Survey-III APOGEE
software postdoc and the NSF grant AST-0807945.  We acknowledge
funding from NSF grants AST-0307851 and AST-0807945, and NASA/JPL
contract 1228235.  We thank the OGLE and MCPS projects for making
their SMC photometric databases available to the public, and the
anonymous referee for useful comments that improved the manuscript.

{\em Facilities:} CTIO (MOSAIC II).


\begin{deluxetable}{lccccccccc}
\tablecaption{MAPS SMC Densities}
\tablecolumns{10}
\tablewidth{0pt}
\setlength{\tabcolsep}{0.05in} 
\tablehead{
\colhead{Field Name\tablenotemark{a}} & \colhead{RA} & \colhead{DEC} & 
\colhead{$R_{\rm SMC}$} & \colhead{$R_{\rm SMC,ellipse}$} & \colhead{PA$_{\rm SMC}$} & 
\colhead{N$_{\rm giants}$} & \colhead{N$_{\rm back}$} & 
\colhead{$\rho_{\rm SMC}$} & \colhead{$\rho_{\rm SMC}$ Error} \\
\colhead{} & \colhead{(J2000)} & \colhead{(J2000)} & 
\colhead{(deg)} & \colhead{(deg)} & \colhead{(deg)} & 
\colhead{ } & \colhead{ } & 
\colhead{stars deg$^{-2}$} & \colhead{stars deg$^{-2}$} 
}
\startdata
190L225a  & 01:09:37.5  & $-$75:05:20.2  & 2.54  & 2.47  & 154.71  & 2581  & 9  & 7,142.6  & 141.4 \\
190L225b  & 01:09:37.7  & $-$74:32:07.8  & 2.08  & 1.93  & 147.21  & 4588  & 9  & 12,718.0  & 188.3 \\
40S026  & 01:12:44.0  & $-$69:10:20.5  & 4.00  & 3.90  & 26.38  & 549  & 8  & 1,502.5  & 65.6 \\
40S071  & 01:39:58.5  & $-$71:10:10.1  & 4.00  & 3.82  & 71.25  & 664  & 8  & 1,819.6  & 72.1 \\
40S116  & 01:46:01.8  & $-$74:11:49.4  & 4.01  & 3.68  & 116.21  & 862  & 9  & 2,367.8  & 82.0 \\
40S161  & 01:15:02.7  & $-$76:33:26.8  & 4.01  & 3.99  & 161.12  & 443  & 10  & 1,202.5  & 59.1 \\
40S206  & 00:23:05.9  & $-$76:18:51.1  & 4.00  & 4.74  & 205.95  & 387  & 10  & 1,046.9  & 55.4 \\
40S251  & 23:58:31.1  & $-$73:41:24.9  & 3.99  & 5.09  & 251.11  & 219  & 9  & 582.8  & 42.0 \\
40S296  & 00:09:14.2  & $-$70:44:37.8  & 3.98  & 4.71  & 296.28  & 258  & 7  & 694.6  & 45.3 \\
40S341  & 00:38:32.0  & $-$69:00:41.0  & 3.99  & 4.13  & 341.40  & 613  & 7  & 1,681.9  & 69.2 \\
51S026  & 01:17:01.9  & $-$68:08:35.3  & 5.10  & 5.08  & 26.31  & 336  & 7  & 911.3  & 51.5 \\
51S116  & 02:01:57.7  & $-$74:24:59.7  & 5.11  & 4.83  & 116.21  & 316  & 9  & 850.7  & 50.1 \\
51S206  & 00:12:06.9  & $-$77:12:56.0  & 5.10  & 5.92  & 206.05  & 94  & 10  & 232.0  & 28.4 \\
51S296  & 23:58:53.2  & $-$70:04:08.2  & 5.08  & 5.87  & 296.29  & 88  & 7  & 223.3  & 27.2 \\
62S004  & 00:56:55.2  & $-$66:38:54.9  & 6.19  & 6.22  & 3.85  & 96  & 7  & 247.0  & 28.2 \\
62S026  & 01:20:59.5  & $-$67:06:38.5  & 6.20  & 6.27  & 26.30  & 84  & 7  & 211.7  & 26.6 \\
190L247a  & 01:45:19.8  & $-$68:40:59.2  & 5.97  & 6.07  & 52.62  & 94  & 8  & 237.2  & 28.1 \\
190L247b  & 01:45:23.8  & $-$68:07:40.3  & 6.41  & 6.56  & 49.45  & 55  & 8  & 129.1  & 22.1 \\
62S116  & 02:18:15.9  & $-$74:33:39.2  & 6.21  & 5.99  & 116.19  & 91  & 9  & 225.5  & 27.9 \\
62S161  & 01:33:01.8  & $-$78:32:00.0  & 6.20  & 6.20  & 161.23  & 117  & 10  & 295.4  & 31.4 \\
62S206  & 23:59:48.7  & $-$78:05:24.3  & 6.19  & 7.10  & 205.98  & 45  & 10  & 94.7  & 20.8 \\
62S229  & 23:33:42.3  & $-$76:08:26.4  & 6.19  & 7.45  & 228.65  & 54  & 10  & 121.2  & 22.3 \\
62S251  & 23:27:09.3  & $-$73:46:09.7  & 6.19  & 7.55  & 251.14  & 23  & 9  & 37.7  & 15.8 \\
62S274  & 23:33:58.2  & $-$71:24:38.8  & 6.18  & 7.38  & 273.69  & 32  & 8  & 65.9  & 17.6 \\
62S296  & 23:48:59.5  & $-$69:21:19.1  & 6.19  & 7.04  & 296.20  & 30  & 7  & 63.3  & 16.9 \\
62S319  & 00:09:27.0  & $-$67:48:38.4  & 6.18  & 6.62  & 318.82  & 59  & 6  & 145.5  & 22.5 \\
62S341  & 00:32:29.6  & $-$66:52:48.6  & 6.19  & 6.34  & 341.28  & 69  & 6  & 173.3  & 24.2 \\
73S026  & 01:24:38.8  & $-$66:04:26.4  & 7.30  & 7.46  & 26.31  & 37  & 7  & 81.4  & 18.6 \\
73S116  & 02:34:51.4  & $-$74:38:19.3  & 7.31  & 7.16  & 116.25  & 36  & 9  & 72.5  & 18.8 \\
73S206  & 23:45:16.0  & $-$78:55:16.4  & 7.29  & 8.29  & 206.05  & 26  & 11  & 40.8  & 17.0 \\
73S296  & 23:40:06.5  & $-$68:36:58.1  & 7.28  & 8.19  & 296.28  & 21  & 6  & 39.3  & 14.7 \\
84S026  & 01:27:56.9  & $-$65:01:32.0  & 8.40  & 8.66  & 26.26  & 27  & 7  & 53.9  & 16.3 \\
84S206  & 23:28:36.8  & $-$79:42:14.4  & 8.39  & 9.49  & 206.07  & 29  & 11  & 47.8  & 17.7 \\
84S251  & 22:56:08.2  & $-$73:33:12.7  & 8.38  & 10.03  & 251.21  & 18  & 9  & 22.9  & 14.6 \\
84S341  & 00:27:34.1  & $-$64:44:27.9  & 8.39  & 8.57  & 341.31  & 37  & 5  & 87.2  & 18.1 \\
95S026  & 01:31:06.9  & $-$63:59:01.2  & 9.50  & 9.87  & 26.31  & 22  & 7  & 40.2  & 15.1 \\
95S206  & 23:09:15.3  & $-$80:25:10.6  & 9.49  & 10.70  & 206.13  & 32  & 12  & 54.8  & 18.5 \\
95S296  & 23:23:41.8  & $-$67:02:55.2  & 9.48  & 10.54  & 296.28  & 19  & 5  & 36.2  & 13.9 \\
106S026  & 01:33:53.8  & $-$62:55:24.9  & 10.60  & 11.09  & 26.23  & 17  & 7  & 26.4  & 13.7 \\
106S206  & 22:47:15.1  & $-$81:04:02.9  & 10.59  & 11.91  & 206.10  & 45  & 12  & 89.4  & 21.1 \\
106S296  & 23:16:16.8  & $-$66:13:51.0  & 10.58  & 11.73  & 296.27  & 15  & 5  & 26.4  & 12.6 \\
\enddata
\tablenotetext{a}{Field names are of format ``radius--L/S--PA'' where radius is 10$\times$ degrees,
L=LMC center, S=SMC center, PA=position angle with respect to L/S.}
\end{deluxetable}

\end{document}